\documentclass[twocolumn,english,reprint,amsmath,amssymb,aps,groupaddress,NOlinenumbers,showpacs]{revtex4-1}
\usepackage{graphicx}
\usepackage{enumerate}
\usepackage{hyperref}
\usepackage{amsthm}
\usepackage{ragged2e}

\newcommand{\Cref}[1]{(\ref{#1})}
\newcommand{\cref}[1]{(\ref{#1})}

\newcommand{\eref}[1]{(\ref{#1})}
\newcommand{\Fref}[1]{Figure~\ref{#1}}
\newcommand{\fref}[1]{Fig.~\ref{#1}}

\usepackage{color}

\begin{document}
\preprint{APS/123-QED}

\title{The driven-dissipative Bose-Hubbard dimer: phase diagram and chaos}

\author{Andrus Giraldo}

\author{Bernd Krauskopf}%
\affiliation{Dodd-Wall Centre for Photonic and Quantum Technologies and Department of Mathematics, The University of Auckland, New Zealand}


\author{Neil G. R. Broderick}
\affiliation{Dodd-Wall Centre for Photonic and Quantum Technologies and Department of Physics, The University of Auckland, New Zealand}%

\author{Juan A. Levenson} 
\author{Alejandro M. Yacomotti}
\affiliation{Centre de Nanosciences et de Nanotechnologies,
  Universit{\'e} Paris-Sud, Universit{\'e}  Paris-Saclay, Palaiseau, France}%

\date{\today}


\begin{abstract}
We present the phase diagram of the mean-field driven-dissipative Bose-Hubbard dimer model. For a dimer with repulsive on-site interactions ($U>0$) and coherent driving we prove that  $\mathbb{Z}_2$-symmetry breaking, via pitchfork bifurcations with sizable extensions of the asymmetric solutions, require a negative tunneling parameter ($J<0$). In addition, we show that the model exhibits deterministic dissipative chaos. The chaotic attractor emerges from a Shilnikov mechanism of a periodic orbit born in a Hopf bifurcation and, depending on its symmetry properties, it is either localized or not.
\end{abstract}

\maketitle


The Bose-Hubbard model is a celebrated fundamental quantum mechanical model that accounts for boson dynamics in a lattice \cite{PhysRevB.40.546}. It successfully describes the interplay between the hopping of particles between neighboring sites of the lattice (with rate $J$) and on-site interactions. 
Such interactions appear as multi-boson terms in the Hamiltonian with interaction energy $U$. Importantly, this model accurately explains the superfluid to Mott insulator phase transitions, that has been experimentally demonstrated in ultracold atomic lattices \cite{Greiner_2002}. A minimal building block in 
this context is the so-called Bose Hubbard dimer, consisting of only two interacting sites, also known as the bosonic Josephson junction \cite{Kellman:2002aa,PhysRevE.74.056608}. Futhermore, the Bose-Hubbard dimer lies at the basis of a number of striking phenomena such as the Josephson effect, self-trapping and symmetry breaking \cite{PhysRevLett.95.010402,PhysRevLett.105.204101,PhysRevA.71.023615}.  

In recent years there has been a growing interest in understanding open quantum systems, where the bosons can be added and destroyed by means of external driving and dissipation mechanisms \cite{RevModPhys.85.299,Schmidt:2013aa,Hartmann_2016,Noh_2016}. In this context, photonic systems have attracted much attention since photons in optical cavities can be injected through an external driving laser, and dissipation comes in as a natural consequence of optical cavity losses. The driven-dissipative Bose-Hubbard dimer has been realized in a number of experimental systems, such as semiconductor microcavities \cite{Abbarchi_2013,PhysRevLett.105.120403}, superconducting circuits \cite{PhysRevX.4.031043} and photonic crystals \cite{Hamel_2015}, where the interactions take the form of Kerr-type optical non-linearities. In many regards, these optical systems constitute outstanding platforms for studying many-body phenomena in open quantum systems \cite{PhysRevX.5.031028}. Among them, dissipative phase transitions \cite{PhysRevX.7.011016,Fink_2017} are an especially exciting open topic, because they provide a conceptual basis for the understanding and the prediction of new collective states, both steady and dynamical ones, with the latter accounting for collective coherent oscillations.

As is well known, phase transitions are characterized by critical phenomena, which can emerge in the thermodynamic limit, i.e., with a large photon number in optical cavities. Remarkably, even single-mode cavities --- i.e., with no spatial degrees of freedom --- may display phase transitions including optical bistability (which is of first order) \cite{PhysRevLett.118.247402,Fink_2017} and the emergence of an oscillation threshold in, e.g., two-photon pumped Kerr resonators \cite{PhysRevA.98.042118} or laser devices \cite{takemura2019low} (which is a second-order phase transition). As pointed out in Ref.~\cite{PhysRevA.98.042118}, driven-dissipative phase transitions are strongly related to mean-field semiclassical solutions in such a thermodynamic limit: these are known as bifurcations of a classical vector field from a non-linear dynamical point of view; see, e.g., \cite{guckenheimer1983nonlinear,Kuz1}. 

Recent examples of complex dynamics found in phase diagrams of driven-dissipative nonlinear cavities include oscillating phases \cite{PhysRevLett.110.163605} and exotic attractors \cite{PhysRevLett.116.143603}. Moreover, Lorenz-type chaos has already been found in the Gross-Pitaevsky equation accounting for incoherent pump of a double potential well Bose-Einstein condensate \cite{PhysRevE.64.025202}. However, to our knowledge, dissipative chaos has not been predicted so far in the coherently driven regime, which is the one that is well described by the driven-dissipative Bose-Hubbard dimer. In this work, we show that dissipative chaos exist in this model and that it is intimately related to Shilnikov homoclinic bifurcations. In addition, we present the phase diagram as a result of the comprehensive analysis of local bifurcations of steady states and their symmetry properties.  

The mean-field approximation of the Bose-Hubbard model is (see, e.g., \cite{Casteels2017}) 
\begin{equation} \label{eq:Coup}
\begin{aligned}
i\dfrac{d \alpha_1}{d \tau} & =  \left( -\Delta - i \dfrac{\gamma}{2}  +2U|\alpha_1|^2 \right) \alpha_1 -J \alpha_2  + F, \\
i\dfrac{d \alpha_2}{d \tau} & =  \left( -\Delta - i \dfrac{\gamma}{2}  +2U|\alpha_2|^2 \right) \alpha_2 - J \alpha_1  + F,  
\end{aligned}
\end{equation}
where $\alpha_1, \alpha_2 \in \mathbb{C}$ are the electric field envelopes in each optical cavity, which are linearly coupled by the tunneling parameter $J$. Both cavities are coherently driven by a field with amplitude $F$ and detuning $\Delta=\omega_p-\omega_c$, where $\omega_p$ and $\omega_c$ are the driving and cavity frequencies, respectively. The cavities have a Kerr-type nonlinearity characterised by $U$, which is assumed to be fast compared to the dissipation rate $\gamma$.  

We apply to system~\cref{eq:Coup} the coordinate transformation 
\begin{eqnarray*} 
(\alpha_1, \alpha_2) \mapsto (A,B) =(-2i \alpha^*_1\sqrt{|U|/\gamma},-2i\alpha^*_2\sqrt{|U|/\gamma}),
\end{eqnarray*} 
where $\alpha^*_i$ is the complex conjugate of $\alpha_i$. With time rescaled to $\tau = \gamma t/2$, 
we obtain the non-dimensionalized Bose-Hubbard model as the vector field
\begin{equation} \label{eq:Coupled} 
\begin{aligned}
\dfrac{d A}{d t} & = -A + i\left(\delta + \text{sign}(U) |A|^2 \right) A +i \kappa B  + f, \\
\dfrac{d B}{d t} & = -B + i\left(\delta   + \text{sign}(U) |B|^2 \right) B+ i \kappa A + f,  
\end{aligned}
\end{equation}
for the (rescaled) envelopes $A, B \in \mathbb{C}$. Here 
\begin{eqnarray*} 
\kappa=-\frac{2 J}{\gamma}, 
f=4F\frac{\sqrt{|U|}}{\gamma^{3/2}}  \ 
{\rm and} \  \delta = -\frac{2\Delta}{\gamma}
\end{eqnarray*} 
are the (rescaled) coupling, drive and detuning parameters, respectively.

Since system~\eref{eq:Coupled} is invariant under the transformation 
\begin{eqnarray*} 
(A,B,U,\delta,\kappa) \mapsto (A^*,B^*,-U,-\delta,-\kappa),
\end{eqnarray*} 
all results for positive $U$ directly translate to those for negative $U$; hence, we consider $U>0$ and  $\text{sign}(U)=1$ in \eref{eq:Coupled} from now. Note that, although $U$ is fixed by the material of the cavity, both the magnitude and sign of $\kappa$ can be changed by means of photonic design in some particular geometries, for example, with potential barrier engineering for the case of photonic crystal nanocavities \cite{Hamel_2015,Haddadi_2014}. Because $\delta$ can be altered at will during the experiment by changing the frequency of the driving laser, states of negative interaction energy in an otherwise positive-$U$ system can be assessed via the parameter transformations $\kappa\mapsto -\kappa$ and $\delta \mapsto -\delta$.

In addition, system~\eref{eq:Coupled} is invariant under the phase-space mirror symmetry of exchanging the two cavities, that is, of swapping $A$ and $B$, which gives rise to $\mathbb{Z}_2$-equivariance of solutions \cite{Kuz1}.  Hence, solutions (equilibria, periodic orbits, trajectories) can be split into two major groups: symmetric and asymmetric ones. For a \emph{symmetric solution} the field intensities and phases in both cavities are the same. The set of all symmetric solutions forms the symmetry subspace given by $A = B$, which is an invariant subspace of system~\eref{eq:Coupled}. An \emph{asymmetric solution}, on the other hand, is one where the intensity and/or the phase in both cavities differ, that is, $A \neq B$. Note that asymmetric solutions often come in pairs, one being the mirror image of the other under swapping $A$ and $B$.

\begin{figure}
\begin{centering}
\includegraphics[clip=true,angle=0,origin=c,width=1\columnwidth]{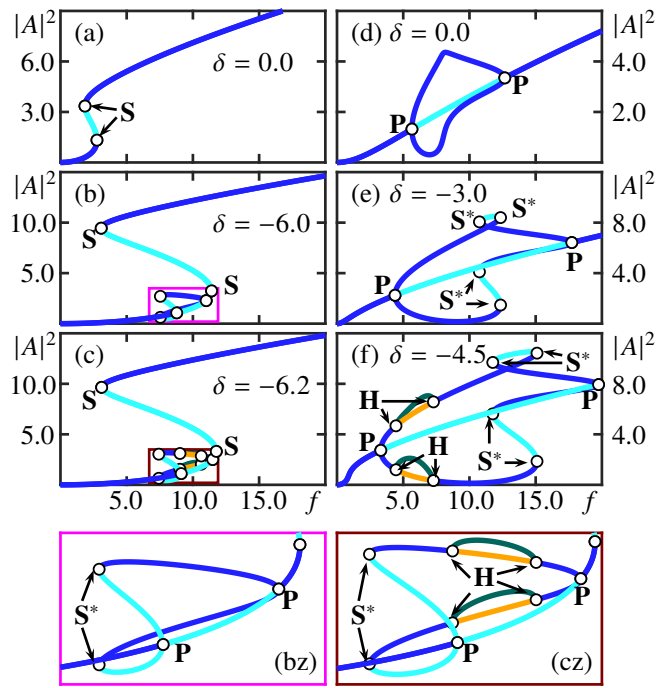}
\par\end{centering}
\centering{}\caption{Phase diagram of system~\eref{eq:Coupled} of intensities $|A|^2$ against the drive parameter $f$ at different $\delta$ values with $\kappa=-3.5$ (left column) and $\kappa=3.5$ (right column); shown are branches of stable equilibria (blue), of saddle equilibria with one unstable eigenvalue (cyan) and with two unstable eigenvalues (orange), and of stable periodic solutions (dark green).} \label{fig:bifSetPlus} 
\end{figure}

We start our analysis of system~\eref{eq:Coupled} by characterizing the equilibria and their bifurcations as the drive parameter $f$ increases. \Fref{fig:bifSetPlus} shows phase diagrams of $|A|^2$ against $f$ for three different values of $\delta$ for the two cases of a negative and positive coupling parameter $\kappa$ in the left and right columns, respectively. The corresponding phase diagrams of $|B|^2$ against $f$ are the same due to $\mathbb{Z}_2$-equivariance. All bifurcations and branches of solutions have been computed with the numerical continuation package \textsc{Auto07p} \cite{Doe2}. When $f = 0$ there is the stable equilibrium given by $A = B = 0$, which gives rise to a monotone branch of stable symmetric equilibria for any $f$ provided the detuning $\delta$ is sufficiently large. However, when $\delta = 0$ and for increasingly negative $\delta$, one finds bifurcations on the branch of symmetric equilibria. 

For negative $\kappa$, as in \fref{fig:bifSetPlus}a-c for $\kappa = -3.5$, we observe for $\delta = 0$ an interval of bistability that is delimited by two saddle-node bifurcations $\mathbf{S}$ of symmetric equilibria [\fref{fig:bifSetPlus}a]. In this interval there are three symmetric equilibria that form a hysteresis loop: a stable equilibrium corresponding to low intensity in both cavities, an intermediate saddle equilibrium, and a stable equilibrium with higher intensity in both cavities. As $\delta$ decreases, the interval of bistability increases in size. While bistability can be encountered in one-cavity systems, spontaneous $\mathbb{Z}_2$-symmetry breaking requires coupled cavities with mirror symmetry; this is the case for the Bose Hubbard model~\cref{eq:Coupled} and spontaneous symmetry breaking phase transitions are known to exist in the form of pitchfork bifurcations \cite{BinMahmud,Casteels2017}. As \fref{fig:bifSetPlus}b for $\delta=-6$ shows, we find two such symmetry breakings at the points labelled $\mathbf{P}$ on the low intensity branch. The left pitchfork bifurcation point $\mathbf{P}$ is subcritical and gives rise to a pair of unstable asymmetric equilibria, while the right point $\mathbf{P}$ is supercritical and gives rise to a pair of stable asymmetric equilibria (one with $|A|^2<|B|^2$ and the other with $|A|^2>|B|^2$); these two branches come together at a pair of saddle-node bifurcations of asymmetric equilibria denoted $\mathbf{S^*}$. These additional bifurcations create a small $f$-interval, between $\mathbf{S^*}$ and the left point $\mathbf{P}$, 
with four stable equilibria: higher and lower intensity symmetric equilibria, and a pair of asymmetric (lower-intensity) equilibria [\fref{fig:bifSetPlus}b]. Notice that between the two points $\mathbf{P}$ there are still three stable objects: the two asymmetric equilibria and the higher-intensity symmetric equilibrium. For an even lower detuning, as in \fref{fig:bifSetPlus}c for $\delta=-6.2$, we find two pairs of supercritical Hopf bifurcations denoted $\mathbf{H}$ on the stable branch of asymmetric equilibria. They give rise to a pair of stable asymmetric periodic orbits, which are represented in the phase diagram by their maxima in $|A|^2$. In the corresponding $f$-interval the asymmetric equilibria are now unstable. Note that the two asymmetric periodic orbits are new attractors that coexist with the stable higher intensity equilibrium.

The situation for $\kappa>0$ is notably different: bistability of the symmetric state is, in fact, absent for all panels \fref{fig:bifSetPlus}d-f. Instead, for $\delta=0$ we find symmetry breaking at two supercritical pitchfork bifurcations, which give rise to a pair of stable asymmetric equilibria in an quite large $f$-interval where the symmetric equilibrium is unstable [\fref{fig:bifSetPlus}d]. This makes this set of abnormal coupling conditions well suited for experimental implementations. Other driving configurations might also allow one to fulfill these conditions, but then the driving phase of the cavities needs to be adjusted properly such that the anti-symmetric state can be linearly excited \cite{Casteels2017}. 

As $\delta$ is decreased, as in \fref{fig:bifSetPlus}e for $\delta=-3.0$, the pair of asymmetric stable equilibria that exists in this interval of symmetry breaking also exhibit saddle-node bifurcations $\mathbf{S^*}$ that create bistability of asymmetric states and a pair of hysteresis loops. Perhaps surprisingly, there is an $f$-range where the intensity in one of the cavities is nearly zero. As \fref{fig:bifSetPlus}f for $\delta=-4.5$ shows, a high negative detuning generates a pair of supercritical Hopf bifurcation points labelled $\mathbf{H}$, which create an $f$-interval with pair of asymmetrical stable periodic solutions, while the asymmetric equilibria are unstable.

\begin{figure}
\begin{centering}
\includegraphics[clip=true,angle=0,origin=c,width=1\columnwidth]{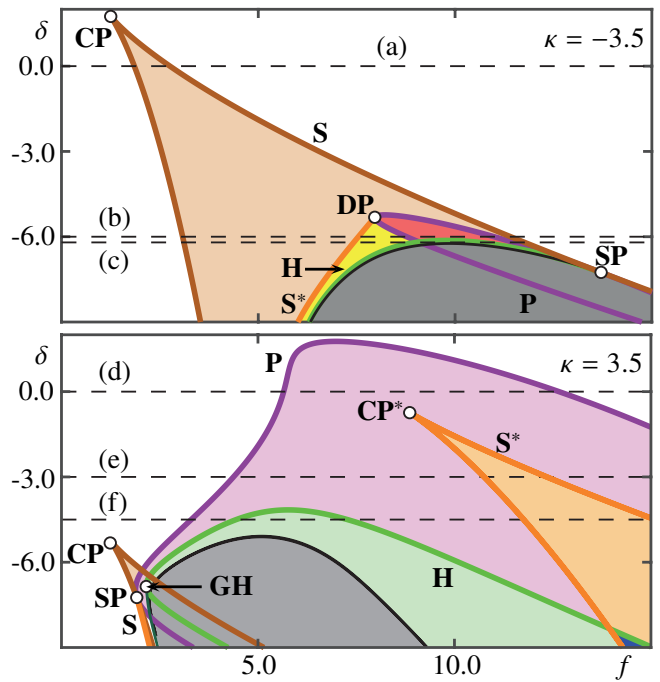}
\par\end{centering}
\centering{}\caption{Phase diagrams of system~\eref{eq:Coupled} in the $(f,\delta)$-plane for $\kappa=-3.5$ (top panel) and $\kappa=3.5$ (bottom panel). Shown are the curve of saddle-node bifurcations of symmetry equilibria $\mathbf{S}$ (brown) and asymmetric equilibria $\mathbf{S^*}$ (orange), of pitchfork bifurcations $\mathbf{P}$ (purple), and of Hopf bifurcations $\mathbf{H}$ (green); also shown are points $\mathbf{CP}$ and $\mathbf{CP^*}$ of cusp, $\mathbf{SP}$ of saddle-node pitchfork, $\mathbf{DP}$ of degenerate pitchfork, and $\mathbf{GH}$ of generalised Hopf bifurcation. Regions are shaded: brown for symmetric bistability; purple for symmetry broken equilibria; orange, yellow and red for multistability of equilibria; green for co-existing stable periodic solutions; and gray for more complex dynamics.} \label{fig:bifSetKappa+2} 
\end{figure}

The bifurcations of equilibria identified in \fref{fig:bifSetPlus} can be followed in the additional parameter $\delta$, yielding a phase diagram in the $(f,\delta)$-parameter plane consisting of bifurcation curves that bound regions with qualitatively different behavior. \Fref{fig:bifSetKappa+2} shows the two phase diagram for both $\kappa=-3.5$ and $\kappa=3.5$ showing the curves $\mathbf{S}$, $\mathbf{S^*}$, $\mathbf{P}$ and $\mathbf{H}$ of saddle-node, pitchfork and Hopf bifurcations, respectively. The dashed horizontal lines correspond to the respective phase diagrams in just $f$ from \fref{fig:bifSetPlus}a-f.

The bifurcation curves $\mathbf{S}$ and $\mathbf{P}$ are drawn from analytical expressions obtained by solving for the equilibria and the respective bifurcation condition \cite{BoHu_longer}. These calculations also show that the saddle-node bifurcation $\mathbf{S}$ in the symmetry subspace (brown line) only occurs if $\delta<-\sqrt{3}-\kappa$ and $1.24081<f$. At the points $(f ,\delta) \approx\left (1.24081, -\sqrt{3} - \kappa\right)$ there is a cusp bifurcation $\mathbf{CP}$ on the curve $\mathbf{S}$ that corresponds to a sharp bound for bistability in the $(f,\delta)$-plane \cite{Kuz1}. The region of bistable symmetric equilibria bounded by $\mathbf{S}$ is shaded brown in both panels of \fref{fig:bifSetKappa+2}; note the quite different position of this region for $\kappa=-3.5$ and for $\kappa=3.5$. Additionally, we find that the curve of pitchfork bifurcation $\mathbf{P}$ (purple curve), which bounds the shaded region with asymmetric equilibria, has a maximum at $\delta = -\sqrt{3}+\kappa$; hence, symmetry breaking only occurs if $\delta< -\sqrt{3}+\kappa$, which is equivalent to the bound found in \cite{MaiaThesis} and also consistent with \cite{Maes:06}. As an unexpected consequence, the condition for symmetry breaking at $\mathbf{P}$ to be possible is that the detuning $\omega_p-(\omega_c-\kappa \gamma/2)$, with respect to the antisymmetric mode, must exceed $\sqrt 3/2$ times the cavity linewidth. This is similar to the bistability condition, but the detuning is measured from the antisymmetric mode of the system, i.e., the one that cannot be linearly excited from the outside world.  These bounds explain why the first phenomenon observed, as $\delta$ decreased, is bistability when $\kappa<0$, while it is symmetry breaking when $\kappa>0$; compare the two panels of \fref{fig:bifSetKappa+2} and see also \fref{fig:bifSetPlus}. Notice that the curve $\mathbf{P}$ is tangent to the curve $\mathbf{S}$ at the saddle-node pitchfork point $\mathbf{SP}$, to the right of the cusp point $\mathbf{CP}$ for $\kappa<0$ and to the left of $\mathbf{CP}$ for $\kappa>0$; moreover, for $\kappa<0$ the curve $\mathbf{P}$ changes criticality at the point $\mathbf{DP}$, from which the curve $\mathbf{S^*}$ of saddle-node bifurcations of asymmetric equilibria emerges.

The curves $\mathbf{SN^*}$ and $\mathbf{H}$ in \fref{fig:bifSetKappa+2} of saddle-node and Hopf bifurcations or asymmetric equilibria are computed numerically by continuation techniques. They bound regions of bistability and of stable asymmetric periodic solutions, respectively. Notice that for $\kappa<0$ the curve $\mathbf{H}$ emanates from a saddle-node pitchfork point $\mathbf{SP}$ and is superctitical throughout, while for $\kappa>0$ we find a change of criticality of $\mathbf{H}$ at a generalised Hopf bifurcation (labelled $\mathbf{GH}$) \cite{Kuz1}.  The considerable amount of multistability of the Bose-Hubbard model~\eref{eq:Coupled} is represented by the overlap between the different shaded region of \fref{fig:bifSetKappa+2}. For even lower negative values of $\delta$ than shown in \fref{fig:bifSetKappa+2}, one can find up to nine equilibria, as pointed out in \cite{BinMahmud}. 

As $\delta$ is decreased for both signs of $\kappa$, the asymmetric stable periodic orbits created at the supercritical Hopf bifurcation $\mathbf{H}$ exhibit bifurcations scenarios to more complex and chaotic dynamics; the region of complex dynamics is indicated by grey shading in \fref{fig:bifSetKappa+2}. Its exact boundary is formed by an intricate arrangement of accumulating bifurcation curves, including period-doubling and different types of homoclinic and heteroclinic bifurcations \cite{BoHu_longer}. Notice that for $\kappa<0$ periodic solutions become more complex almost immediately, while for $\kappa>0$ we find a quite large region of stable periodic solutions. While a full discussion of the different scenarios is beyond the scope of this paper, we conclude by showing the existence of two types of Shilnikov-type chaos in the Bose-Hubbard model: asymmetric chaotic dynamics localized in one of the cavities, and symmetric chaotic dynamics with irregular switching between the two cavities. 

\begin{figure}
\begin{centering}
\includegraphics[clip=true,angle=0,origin=c,width=1\columnwidth]{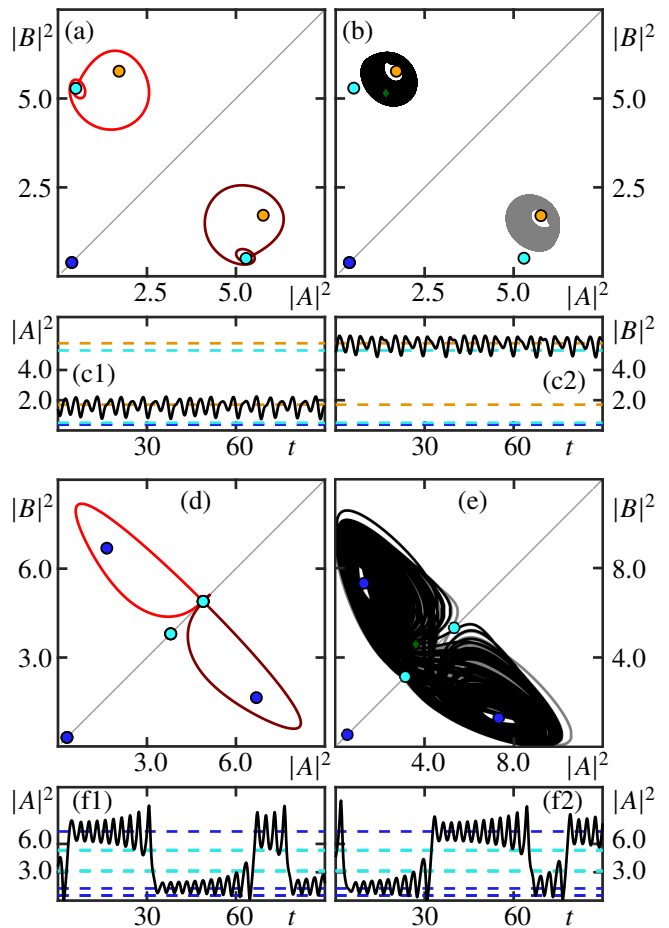}
\par\end{centering}
\centering{}\caption{Shilnikov orbits (a,d), trajectories on chaotic attractors (b,e) in the $(|A|^2, |B|^2)$-plane, and temporal traces of  $|A|^2$ (c1,f1)  and of $|B|^2$ (c2,f2) for the black trajectories. Here $\kappa = -3.5$, $\delta = -8$ with $f \approx 6.943$ for~(a) and $f = 6.93$ for~(b,c); and $\kappa = 3.5$, $\delta = -8$ with $f \approx 2.379$ for~(d) and $f = 3.0$ for~(e,f).} \label{fig:hom} 
\end{figure}

At a homoclinic bifurcation a system of differential equations posesses a special trajectory, called a homoclinic orbit, that converges both in backward and forward time to a saddle equilibrium. For the special case that the equilibrium is a saddle focus (with a pair of complex conjugate eigenvalues),  a celebrated result by Shilnikov \cite{Shil5} states that (under a certain eigenvalue conditions) infinitely many periodic orbits and chaotic behavior can be found nearby. This phenomenon is now referred to as a Shilnikov bifurcation. In the $\mathbb{Z}_2$-equivariant system~\eref{eq:Coupled} we find Shilnikov bifurcations both to asymmetric saddle foci and to symmetric saddle focus, as well as two types of nearby chaotic attractors.

\Fref{fig:hom} shows both cases of an asymmetric and of a symmetric Shilnikov bifurcation and associated chaotic attractors. For $\kappa<0$ in \fref{fig:hom}a, there exist asymmetric Shilnikov homoclinic orbits to  each of the pair of asymmetric equilibria, one either side of the symmetry line $|A|^2 = |B|^2$. For nearby $f$ as in \fref{fig:hom}b, we find a pair of chaotic asymmetric and localized attractors. The time traces in panels (c1,c2) show that this type of localized chaotic dynamics is characterized by a dominance in intensity of one of the two cavities for all time. For $\kappa>0$, on the other hand, we find a pair of Shilnikov homoclinic orbits to a single symmetric saddle equilibrium [\fref{fig:hom}d]. This has the consequence that the associated chaotic attractor in \fref{fig:hom}e is much larger and no longer localized as it crosses the line $|A|^2 = |B|^2$. As the time traces in panels (f1,f2) clearly show, this type of chaotic dynamics is characterized by episodes of dominance in intensity of one of the two cavities for some time, with rapid and irregular switching events of the intensity to the other cavity. We remark that a local analysis of equilibria alone is not able to predict and explain this global switching behavior of the system. 

In conclusion, our systematic bifurcation analysis produced a comprehensive phase diagram description of both equilibria and bifurcating non-equilibrium solutions of the driven-dissipative Bose-Hubbard dimer model. In particular, we were able to identify bifurcations to periodic solutions, and subsequent localized and non-localized chaotic behavior near homoclinic bifurcations of Shilnikov type. 
Crucial to the dynamics is the fact that the two cavities are coherently driven, which means that the system remains four-dimensional since there is no phase freedom as in active cavities. Interestingly, in a system with positive on-site interaction energy ($U>0$), non-localized Shilnikov chaos can be observed for negative tunneling parameter ($J<0$), in a bifurcation cascade whose details will be reported elsewhere. Our mean-field dynamical results thus pave the way for studying complex non-equilibrium orbits --- including quantum dissipative chaos --- in the quantum master equation of the open Bose-Hubbard dimer. 

This work has been partially funded by the \lq \lq Investissements d'Avenir" program (Labex NanoSaclay, Grant No. ANR-10-LABX-0035) and the ANR UNIQ DS078.


%

\end{document}